\documentclass[12pt]{article}
\usepackage{amsmath,amssymb}
\newcommand{\be}{\begin{equation}}
\newcommand{\bea}{\begin{eqnarray}}
\newcommand{\ena}{\end{eqnarray}}
\newcommand{\nb}{\nonumber}
\newcommand{\e}{{\rm e}}

\newcommand{\F}{{\cal F}}

\newcommand{\Si}{{\cal S}}
\newcommand{\M}{{\cal M}}
\newcommand{\A}{{\cal A}}

\newcommand{\bx}{{\bf x}}
\newcommand{\bp}{{\bf p}}
\newcommand{\der}{\partial }
\newcommand{\simp}{\left ( \sigma ,\sigma_b\right )}
\newcommand{\alg}{\A_{(\sigma,\sigma_b)}}
\newcommand{\Sr}{{\cal S}_{\rm ren}}

\newcommand{\RR}{\mbox{${\mathbb R}$}}

\begin{document}

\thispagestyle{empty}

\rightline {IFUP-TH 13/2001}

\vskip 1 truecm
\centerline {\Large \bf Local Fields on the Brane Induced by}
\bigskip
\centerline {\Large \bf Nonlocal Fields in the Bulk}
\vskip 1.5 truecm
\centerline {\large \rm Mihail Mintchev}
\medskip
\bigskip
\centerline {\it Istituto Nazionale di Fisica Nucleare, Sezione di Pisa}
\centerline {\it Dipartimento di Fisica dell'Universit\`a di Pisa,}
\centerline {\it Via Buonarroti 2, 56127 Pisa, Italy}
\bigskip
\medskip

\vskip 3.5 truecm
\centerline {\large \it Abstract}
\medskip
\bigskip
We investigate quantum field theory in a bulk space with boundary, 
which represents
a $3$-brane. Both flat and anti-de Sitter backgrounds are considered.
The basic idea is to keep local commutativity only on
the brane, giving up this requirement in the bulk. We explore the consequences
of this proposal, constructing a large family of nonlocal bulk 
fields, whose brane
relatives are local. We estimate the ultraviolet behavior of these local
brane fields, characterizing a subfamily which
generates renormalizable theories on the brane. The issue of brane 
conformal invariance
and the relation between bulk and brane conserved currents are also examined in
this framework.

\bigskip

\bigskip
\bigskip
\bigskip

\centerline {March 2001}
\vfill \eject

\section{\bf Introduction}

\bigskip

Triggered by the fast development of brane theory (cf.\cite{P} and 
references therein)
and the revival \cite{ADD1,ADD2,AADD,RS1,RS2} of the attractive idea 
\cite{RS} to
consider the observable universe as a three-brane embedded in a space with
more (non)compact dimensions, there is recently great interest in quantum field
theory with extra space dimensions. The intriguing fact that strongly motivates
the investigation of such a scenario is the possibility \cite{GRW} to test it
experimentally at high-energy colliders. In the context of 
electroweak interactions,
extra dimensions provide new mechanisms for symmetry breaking 
\cite{A, AHNSW} and
the generation of fermion mass hierarchies \cite{ADD1,AADD,AS,RS1}.
Despite of the great progress in the subject, some
relevant phenomenological and theoretical questions are still open.
Among others, we have in mind the following general problem.
Let $\{\Phi_i\}$ be a system of quantum fields  propagating in a bulk 
space $\M$
with  extra dimensions and with a nontrivial boundary $\der\M$, 
representing our 3+1
dimensional space-time (3-brane). The problem is to characterize
the fields $\{\varphi_i\}$ induced by $\{\Phi_i\}$ on $\der\M$.
It is natural to expect (\cite{tH,S}) that the
correspondence $\{\varphi_i\} \leftrightarrow \{\Phi_i\}$
represents an essential point in understanding the deep relation
between bulk and brane dynamics.
Since in passing from $\M$ to $\der\M$ one is suppressing spatial 
dimensions, it is
not surprising that local fields in the bulk give rise to local fields on the
brane. For a complete understanding of the map
$\{\varphi_i\} \leftrightarrow \{\Phi_i\}$ however, it is essential to
know also whether there
exist bulk fields, which in spite of being nonlocal in $\M$, induce
local fields on the brane $\der\M$. The physical
relevance of this question stems from the observation that local commutativity,
which is a crucial
prerequisite on the brane, in principle may be violated in the bulk.
The main goal of the present paper is to explore this possibility.
We explicitly construct a class of nonlocal bulk fields $\{\Phi_i\}$ which
induce local fields $\{\varphi_i\}$ on $\der\M$, respecting all 
fundamental physical
requirements like Poincar\'e invariance, spectral condition and positivity.
We first investigate the characteristic features of $\{\varphi_i\}$, focusing
afterwards on two distinguished subfamilies which generate 
renormalizable and conformal
invariant brane theories respectively.
We analyze also some related aspects concerning the interplay between 
bulk and brane
symmetries, studying in particular brane vector currents induced by 
conserved currents
in the bulk. We start by considering a flat bulk space $\M$ and then 
discuss the
case when $\M$ is a slice of anti-de-Sitter (AdS) space-time, which attracts
much attention \cite{V,ANO,APR,RZ} in relation with the 
Randall-Sundrum proposal
\cite{RS1,RS2} for brane localization of gravity.

The paper is organized as follows.
In the next section we describe the framework and construct
a specific class of bulk fields and their brane relatives. We examine
in detail the influence of the brane on the quantization, detecting 
and parametrizing the
freedom left after imposing the boundary conditions.
The central points are the analysis of locality and the issues of 
renormalizability
and conformal invariance. In section 3 we establish a bridge between bulk and
brane symmetries. Here we derive also Ward identities on the brane.
Section 3 extends the formalism to an AdS background. The last section is
devoted to our conclusions.

\bigskip

\section{\bf Quantization in the presence of a brane}

The purpose of this section is to develop a general framework for studying
quantum fields induced on branes. For a bulk space we take the manifold
$\M = \RR^{4} \times \RR_+$,  where $\RR_+$ is the half line $\{y\in 
\RR \, :\, y>0 \}$.
It is convenient to adopt  the coordinates
$(x,y) \in \RR^{4} \times \RR_+$ where $x \equiv (x^0,x^1,x^2,x^3) = 
(x^0,\bx)$.
In this section $\M$ is equipped with the flat metric
\be
G_{\alpha \beta } =
\begin{pmatrix}g & 0 \\ 0 & -1\end{pmatrix}\quad ,
\qquad {\rm diag}\, g = (1,-1,-1,-1) \, , \qquad \alpha , \beta = 0,...,4 \, .
\label{met}
\end{equation}
The boundary $\der M$ coincides with the $3+1$-dimensional Minkowski space
${\bf M}_{3+1} \equiv \{{\RR}^{4},\, g\}$, representing the $3$-brane.

In order to illustrate in its simplest form the general mechanism for inducing
quantum fields on ${\bf M}_{3+1}$ from $\M = \RR^{4} \times \RR_+$, we
consider a free Hermitian scalar field with mass $M\geq 0$. The 
corresponding action reads
\be
S_0 = \frac{1}{2} \int_{-\infty}^\infty d^4x  \int_0^\infty dy
:\left ( \der^\alpha \Phi \der_\alpha \Phi - M^2 \Phi \Phi \right ): (x,y)
  - \frac{\eta }{2}\int_{-\infty}^\infty d^4x :\Phi \Phi : (x,0) \, ,
\label{act}
\end{equation}
where $\eta \in \RR $ and $: ... :$ indicates the normal product with respect
to the creation and annihilation operators introduced few lines below.
The variation of $S_0$ gives both the equation of motion
\be
\left (\der^\alpha \der_\alpha + M^2 \right ) \Phi (x,y) =  0 \, ,
\label{eqm}
\end{equation}
and the so called mixed boundary condition
\be
\lim_{y \downarrow 0} \left (\der_y - \eta \right )\Phi (x,y) = 0 \, .
\label{bcphi}
\end{equation}
The parameter $\eta $ has dimension of a mass; in the limits
$\eta \to 0$ and $\eta \to \infty $ one recovers from (\ref{bcphi}) 
the Neumann and
Dirichlet boundary conditions respectively. Other boundary conditions
\cite{GKV} can be treated analogously.

We quantize eqs.(\ref{eqm},\ref{bcphi}) using the operator formalism, 
which allows a
better control on locality and positivity then the functional 
integral approach.
As already explained in \cite{MP},
stability considerations imply the lower bound
\be
\eta \geq -M \, ,
\label{stab}
\end{equation}
which is assumed throughout the paper. The field $\Phi$, satisfying
eqs.(\ref{eqm},\ref{bcphi}),  admits the decomposition
\bea
\Phi (x,y) = \int_{-\infty}^\infty \frac{d^3p}{(2\pi)^3}
\Biggl \{
b^*(\bp)E_{M^2 - \eta^2}(x, \bp ) \psi_b (y) +
b(\bp){\overline E}_{M^2 - \eta^2} (x, \bp ) {\overline \psi}_b(y) + \nb \\
\int_0^\infty \frac{d\lambda}{2\pi}
\left[a^*(\bp, \lambda) E_{M^2 + \lambda^2}(x, \bp ) \psi (y,\lambda) +
a(\bp, \lambda ) {\overline E}_{M^2 + \lambda^2}(x, \bp )
{\overline \psi}(y,\lambda)\right ] \Biggr \} \, ,
\label{f1}
\ena
where
\be
E_{m^2} (x, \bp ) = \frac{1} {\sqrt {2\omega_{m^2}(\bp )}}\,
\e^{-i\omega_{m^2}(\bp )x^0+i\bp \bx } \, , \qquad
\omega_{m^2}(\bp ) = \sqrt {\bp^2 + m^2} \, ,
\label{xw}
\end{equation}
\be
\psi (y,\lambda)  = \e^{-i\lambda y} +
\frac{\lambda - i\eta}{\lambda + i\eta}\, \e^{i\lambda y}\, , \qquad
\psi_b (y) = \theta (-\eta )\sqrt {2|\eta|}\, \e^{\eta y} \, ,
\label{wy}
\end{equation}
and the bar stands for complex conjugation. The basic ingredients
of the superposition (\ref{f1}) are essentially two:

\begin{description}

\item {(i)} the system of eigenfunctions (\ref{xw},\ref{wy});

\item {(ii)} the set $\{a^*(\bp,\lambda ),\, a(\bp,\lambda ),\, 
b^*(\bp),\, b(\bp)\, :
\, \bp \in \RR^3\, ,\lambda \in \RR_+ \}$ of creation and 
annihilation operators.

\end{description}

\noindent The functional input (i) is uniquely determined by the 
equation of motion
(\ref{eqm}) and the boundary condition (\ref{bcphi}). Besides the familiar
plane waves $E_{m^2}(x,\bp )$, Eq.(\ref{f1})  involves the functions (\ref{wy})
related to the Hamiltonian operator $-\der_y^2$ on $\RR_+$:
$\psi (y, \lambda)$ describe scattering
states, whereas $\psi_b (y)$ is the unique bound state (with energy 
$- \eta^2 $)
present only for $\eta < 0$. These functions form an orthonormal system and
satisfy the completeness relation
\be
\int_0^\infty \frac{d\lambda}{2\pi} \,{\overline \psi}(y_1,\lambda )
{\psi}(y_2,\lambda) +
\theta (-\eta ) \psi_b(y_1) \psi_b(y_2) =
\delta (y_1-y_2) \, , \qquad y_1,\, y_2 \in \RR_+ \, ,
\label{compl}
\end{equation}
with
\be
\theta (\alpha ) = \left\{ \begin{array}{cc} 1 & \mbox{if } \alpha > 0\, , \\
0 & \mbox{if } \alpha \leq 0 \, .
\end{array} \right.
\label{theta}
\end{equation}

Concerning the algebraic input (ii), there is certain freedom reflecting
an intrinsic feature of the system under consideration, namely, the breakdown
of bulk Poincar\'e invariance due to the brane. This freedom can be 
parametrized
by a pair $\simp $, where $\sigma (\lambda )$ is a distribution
on $[0,\infty )$ and $\sigma_b \in \RR$. Any $\simp $ defines an algebra
$\alg $ generated by $\{a^*(\bp,\lambda ),\, a(\bp,\lambda ),\, 
b^*(\bp),\, b(\bp)\}$
subject to the constraints:
\bea
&&[a(\bp_1, \lambda_1)\, ,\, a^*(\bp_2, \lambda_2)] =
(2\pi)^4\, \delta (\bp_1-\bp_2)\, \sigma(\lambda_1 )\, \delta 
(\lambda_1-\lambda_2)\, ,
\label{ccr1} \\
&&[a(\bp_1, \lambda_1)\, ,\, a(\bp_2, \lambda_2)] =
[a^*(\bp_1, \lambda_1)\, ,\, a^*(\bp_2, \lambda_2)] = 0 \, , \label{ccr0}\\
&&[b(\bp_1)\, ,\, b^*(\bp_2)] =
(2\pi)^3\, \sigma_b\, \delta (\bp_1-\bp_2) \, ,
\label{ccr2} \\
&&[b(\bp_1)\, ,\, b(\bp_2)] = [b^*(\bp_1)\, ,\, b^*(\bp_2)] = 0 \, . 
\label{ccr00}
\ena
The requirement of positivity of the
metric in the Fock representations $\F(\alg)$ of $\alg $ implies the 
restrictions
\be
\sigma (\lambda )\geq 0\, , \qquad \sigma_b \geq 0 \, .
\label{pos}
\end{equation}

The data $\simp $ have a deep structural impact
on different levels. We focus first on the equal-time $\Phi$-commutators
generated by (\ref{ccr1}, \ref{ccr2}). The nontrivial one is
\bea
\left [(\der_0\Phi )(x^0,\bx_1,y_1)\, ,\, \Phi (x^0,\bx_2,y_2)\right ] =
\qquad \qquad \qquad \nb \\
-i\delta (\bx_1-\bx_2) \left [
\int_0^\infty \frac{d\lambda}{2\pi} \,{\overline \psi}(y_1,\lambda ) 
\sigma (\lambda )
{\psi}(y_2,\lambda) +
\theta (-\eta ) \sigma_b \psi_b(y_1) \psi_b(y_2) \right ]
\label{ccr3}
\ena
and represents a generalization of the standard canonical commutation relation.
Because of (\ref{compl}), one recovers the latter for $\simp = (1,1)$.
We therefore conclude that the pair $\simp $ parametrizes
a class $\Si$ of generalized canonical structures. Each of them 
defines an initial
condition for the time evolution in the bulk, which is compatible with the
bulk symmetries. It will be shown later on, that
besides this bulk interpretation, $\simp$ has a clear physical meaning
also from the brane view point.

Let us investigate now the locality properties of $\Phi $. One easily derives
\bea
\left [\Phi (x_1,y_1)\, ,\, \Phi(x_2,y_2) \right ] =
-2i\, \sigma_b\, \theta (-\eta )|\eta|\e^{\eta {\widetilde y}_{12}}\,
D_{M^2-\eta^2}(x_{12}) +
\qquad \qquad \nb \\
-i \int_0^\infty \frac{d\lambda}{2\pi}\, \frac{4\sigma(\lambda 
)}{\lambda^2+\eta^2}\,
\left (\lambda \cos\lambda y_1 + \eta \sin \lambda y_1 \right )
\left (\lambda \cos\lambda y_2 + \eta \sin \lambda y_2 \right )
D_{M^2+\lambda^2}(x_{12}) \, ,
\label{ppc}
\ena
where $x_{12} = x_1-x_2$, ${\widetilde y}_{12} = y_1+y_2$ and
\be
D_{m^2}(x)
= i \int_{-\infty}^\infty \frac{d^4p}{(2\pi)^4} \,
\e^{-ipx} [\theta (p^0) - \theta (-p^0)]\, 2\pi \delta (p^2 - m^2) \, ,
\qquad p=(p^0,\bp)\, ,
\label{PJ}
\end{equation}
is the well-known Pauli-Jordan function in ${\bf M}_{3+1}$.
In (\ref{ppc}) and in what follows, the integration in $\lambda$ is
understood in the sense of distributions, i.e. after smearing with
test functions in $(x_1, y_1)$ and $(x_2, y_2)$. One can demonstrate \cite{MP}
that $\Phi $ is a local field for $\simp = (1,1)$. We observe in passing that
this property is not so straightforward as on manifolds without 
boundaries, because
one must take into account \cite{LMZ,LM} that signals propagating in 
the bulk are
reflected from the brane as well. For generic $\simp \not= (1,1)$
the field $\Phi $ is nonlocal, i.e. there exist space-like separated points
$(x_1,y_1)$ and $(x_2,y_2)$ in which the right hand side of eq.(\ref{ppc}) does
not vanish. Nevertheless, we will prove below that local commutativity
is restored on the brane.

We have shown so far that even for a fixed boundary condition 
(\ref{bcphi}), the
quantization of the action $S_0$ is not unique if the requirement of locality
in the bulk is relaxed.
Each pair $\simp \in \Si$ defines by means of 
eqs.(\ref{f1},\ref{ccr1},\ref{ccr2}) a
bulk field $\Phi $, which is in general nonlocal, though
generated by a local bulk action. As already mentioned, the origin of
this unusual feature is the defect, produced in the bulk by the brane.
Being codified in the algebra $\alg$, this is a genuine quantum phenomenon.

The two-point vacuum expectation value (Wightman function) of
$\Phi $ in the Fock space $\F(\alg)$ reads
\bea
\langle \Phi (x_1,y_1) \Phi(x_2,y_2) \rangle_{{}_0} =
2\, \sigma_b\, \theta (-\eta )|\eta|\e^{\eta {\widetilde y}_{12}}\,
W_{M^2-\eta^2}(x_{12}) + \qquad \qquad \nb \\
\int_0^\infty \frac{d\lambda}{2\pi}\, \frac{4\sigma(\lambda 
)}{\lambda^2+\eta^2}\,
\left (\lambda \cos\lambda y_1 + \eta \sin \lambda y_1 \right )
\left (\lambda \cos\lambda y_2 + \eta \sin \lambda y_2 \right )
W_{M^2+\lambda^2}(x_{12}) \, ,
\label{ppb}
\ena
where
\be
W_{m^2}(x)
= \int_{-\infty}^\infty \frac{d^4p}{(2\pi)^4} \,
\e^{-ipx} \theta (p^0) 2\pi \delta (p^2 - m^2) \,
\label{w}
\end{equation}
is the two-point scalar function of mass $m^2$ in ${\bf M}_{3+1}$. 
Since $\Phi$ is
a free field, eq.(\ref{ppb}) completely fixes all of its correlation functions.
The latter define via
\be
\langle \varphi (x_1)\cdots \varphi (x_{n})
\rangle_{{}_0} =
\lim_{y_i \downarrow 0}\, \langle \Phi (x_1,y_1)\cdots \Phi 
(x_{n},y_{n}) \rangle_{{}_0} \, ,
\label{corg}
\end{equation}
the field $\varphi $ induced on the brane.
The existence of the limit follows directly from eq. (\ref{ppb}).
The fundamental features
of $\varphi $ are encoded in the two-point function
\bea
w(x_{12}) = \langle \varphi (x_1)\varphi (x_2)\rangle_{{}_0} =
\qquad \qquad \qquad \qquad \nb \\
\int_0^\infty \frac{d\lambda}{2\pi}\, \frac{4\lambda^2 
\sigma(\lambda)}{\lambda^2+\eta^2}\,
W_{M^2+\lambda^2}(x_{12}) +
2\sigma_b\, \theta (-\eta )|\eta|\, W_{M^2-\eta^2}(x_{12}) \, .
\label{bpp0}
\ena
An obvious change of variables leads to the K\"all\'en-Lehmann representation
\bea
w(x_{12}) =
\int_0^\infty d\lambda^2 \, \varrho (\lambda^2 )\, W_{\lambda^2}(x_{12}) \, ,
\label{bpp}
\ena
with
\be
\varrho (\lambda^2 ) = \theta (\lambda^2 - M^2)
\frac{\sqrt {\lambda^2-M^2}\, \sigma (\sqrt {\lambda^2 -M^2}\, )}
{\pi(\lambda^2+\eta^2-M^2)} +
2\, \sigma_b\, \theta (-\eta )|\eta| \delta (M^2-\eta^2-\lambda^2) \, ,
\label{rotot}
\end{equation}
Therefore, $\varphi $ is a generalized free field, which can be fully
reconstructed (see e.g. sect. II.6 of \cite{J}) from the two-point function
(\ref{bpp0}). The conditions (\ref{pos}) imply that
$\varrho (\lambda^2 )d\lambda^2$ is
a positive measure on $[M^2,\, \infty )$, thus ensuring that the underlying
state space is a Hilbert space.  Brane Poincar\'e invariance is manifest.

The spectrum of the $\varphi$-mass operator
belongs to $[M^2,\, \infty)$ and if $-M \leq \eta <0$
has an additional pure point contribution at $M^2-\eta^2$. The states
contributing to $[M^2,\, \infty)$ represent the Kaluza-Klein (KK) modes,
whereas $M^2-\eta^2$ is the mass of the state corresponding to $\psi_b$.
In this context the pair $\sigma $ defines the weight by which
each KK mode contributes to the field $\varphi $. This is actually the physical
interpretation of $\sigma $ on the brane level. For
$\eta = 0$ and $\sigma (\lambda ) = \pi \mu \delta (\lambda^2-\mu^2)$,
the free field with mass $m^2 = M^2+\mu^2$ is localized on the brane.

The evaluation of the integral in Eq. (\ref{bpp}) is straightforward in
momentum space. For the Fourier transform ${\widehat w}$ of $w$ one gets
\be
{\widehat w}(p) = 2\theta (p^0) \left [ \theta (p^2-M^2)
\frac{\sqrt {p^2-M^2}\, \sigma (\sqrt {p^2 -M^2}\, )}
{p^2-M^2 + \eta^2} + \sigma_b\, \theta (-\eta) |\eta |2\pi
\delta(p^2 -M^2 + \eta^2) \right ] \, ,
\label{ft1}
\end{equation}
implying that for generic $\sigma$ the induced field $\varphi $ does 
not satisfy the free
Klein-Gordon equation and cannot be derived from a local action 
(integral of a local density)
on the brane. Nevertheless, $\varphi $ is a local field because
\be
\left [\varphi (x_1)\, ,\, \varphi (x_2)\right ] =
-i \int_0^\infty d\lambda^2 \, \varrho (\lambda^2 )\, 
D_{\lambda^2}(x_{12}) \, .
\label{bppc}
\end{equation}
Thus the limit $y\to 0$ absorbs all
noncausal effects, regarding the behavior of $\Phi $. From 
(\ref{bppc}) one gets
\be
\left [(\der_0\varphi) (x^0,\bx_1)\, ,\, \varphi (x^0,\bx_2)\right ] =
-i\, \delta (\bx_1-\bx_2)\, \int_0^\infty d\lambda^2 \, \varrho 
(\lambda^2 )\, .
\label{ppcc}
\end{equation}
Therefore, if
\be
C \equiv \int_0^\infty d\lambda^2 \, \varrho (\lambda^2 ) < \infty \, ,
\label{ffr}
\end{equation}
$\varphi $ satisfies the conventional canonical commutation relation up to
a finite multiplicative field renormalization.

The bound (\ref{ffr}) selects
a subclass $\Sr \subset \Si$ of pairs $\simp $, generating
brane fields with distinguished ultraviolet (UV) behavior.
Indeed, from (\ref{bppc}) one obtains for the propagator
\be
\tau (x_{12}) \, = \,i \langle T \varphi (x_1)\varphi (x_2)\rangle_{{}_0} =
\int_0^\infty d\lambda^2 \, \varrho (\lambda^2 )\, 
\Delta_{\lambda^2}(x_{12}) \, ,
\label{tx}
\end{equation}
$T$ indicating time ordering and
\be
\Delta_{m^2}(x) \, = \, - \int_{-\infty}^\infty
\frac{d^4p}{(2\pi)^4} \,
\frac{\e^{-ipx}}{p^2 - m^2 + i\varepsilon }  \, .
\label{tf}
\end{equation}
In momentum space one has
\be
{\widehat \tau } (p) \, = \, - \int_0^\infty d\lambda^2 \, \varrho 
(\lambda^2 )\,
\frac{1}{p^2 - \lambda^2 + i\varepsilon } \, .
\label{tp}
\end{equation}
Combining (\ref{ffr}) with (\ref{tp}), one gets the estimate
\be
p_{{}_E}^2\, {\widehat \tau } (p_{{}_E}) \leq C \, , \qquad p_{{}_E} 
= (-ip^0, \bp) \, .
\label{est1}
\end{equation}
Therefore, at large Euclidean momenta $p_{{}_E}$ the
$\varphi$-propagator decays at least like $1/{p_{{}_E}^2}$, provided 
that (\ref{ffr})
is satisfied. This is not the case when (\ref{ffr}) is violated. In 
fact, for the local
(canonical) bulk field  $\simp = (1,1)$, considered till now in the literature,
one finds
\be
{\widehat \tau }_{{}_{(1,1)}}(p) =
  - \frac{\sqrt{M^2 - p^2} - \eta}{p^2 -M^2 + \eta^2 + i \epsilon } \, ,
\label{tpc}
\end{equation}
which decays like $1/{\sqrt {p_{{}_E}^2}}$ when $p_{{}_E}^2 \to \infty$.

The above result can be used to construct renormalizable theories on the brane.
The simplest example coming in mind is the model
\be
S = S_0 + \frac{g}{3!} \int_{-\infty}^\infty d^4x :\Phi^3: (x,0) \, .
\label{int}
\end{equation}
Giving up local commutativity in the bulk implies that the
quantization of (\ref{int}) is not uniquely defined: one must choose
$\simp \in \Si$. If $\simp \in \Sr$, one gets both a renormalizable
and local theory on the brane.
A concrete example is $\simp = \left (\mu^2/\lambda^2, \sigma_b\right )$,
where $\mu$ is a parameter. Evaluating the integral (\ref{tp}), one gets
\be
{\widehat \tau }_{{}_{(\mu^2/\lambda^2, \sigma_b)}}(p) =
  - \frac{1}{p^2 -M^2 + \eta^2 + i \epsilon }
\left [ \frac{\mu^2\sqrt{M^2 - p^2}}{p^2 -M^2 + i \epsilon } +
\frac{\mu^2}{|\eta|} + 2 \sigma_b\, \theta (-\eta) |\eta| \right ] \, ,
\label{renp}
\end{equation}
leading to a renormalizable perturbative expansion for the
correlation functions of $\varphi$.

One may wonder if by a suitable choice of $\simp$ one cannot further 
improve the
$1/{p_{{}_E}^2}$ UV-decay of (\ref{renp}), getting a perturbatively finite
theory on the brane. We observe in this respect that the restrictions 
(\ref{pos})
imply the estimate
\be
{\widehat \tau } (p_{{}_E}) \geq \frac {C^\prime (\Lambda 
)}{p_{{}_E}^2 + \Lambda^2}\, ,
\label{est2}
\end{equation}
$\Lambda$ and $C^\prime (\Lambda )$ being some constants. Therefore, positivity
of the metric in the state space prevents an UV-decay faster then 
$1/{p_{{}_E}^2}$.
On the contrary, allowing for violations of (\ref{pos}), one can 
obtain UV-finite
models on the brane. For instance, setting
\be
\sigma_b = -\frac{\mu^2}{2\eta^2}\, , \qquad \eta < 0 \, ,
\label{fin}
\end{equation}
in eq.(\ref{renp}), one has UV-finite perturbative expansion for the
model (\ref{int}). This property can be used for developing an 
UV-regularization
procedure on the brane.

We conclude this section with a discussion of brane conformal invariance.
Let $M = \eta = 0$ and let us consider the one-parameter family
\be
\sigma (\lambda ) = \frac {\pi}{\Gamma (d-1)}\, \lambda^{2d-3} \, ,
\label{confden}
\end{equation}
postponing for a while the justification of the normalization factor. We also
require that $\sigma (\lambda )$ has at most integrable singularity 
in $\lambda = 0$,
namely
\be
d>1 \, .
\label{d}
\end{equation}
Since $\eta=0$, the value of $\sigma_b$ is irrelevant and inserting
(\ref{confden}) in  eq.(\ref{ft1}), one finds
\be
{\widehat w}(p) = \frac{2\pi}{\Gamma(d-1)}\, \theta (p^0) \theta 
(p^2) (p^2)^{d-2}
\equiv \frac{2\pi}{\Gamma(d-1)}\, \theta (p^0) (p^2)_+^{d-2} \, ,
\label{conf}
\end{equation}
which is (see e.g. sect.IV.1 of \cite{TMP}) the two-point function of a
conformal covariant scalar field
of dimension $d$, the lower bound (\ref{d}) ensuring positivity. Therefore,
our framework provides a mechanism for generating local conformal 
scalar fields on the brane
of any dimension $d>1$. With the exception of $d=\frac{3}{2}$, all these fields
are induced by nonlocal bulk fields, which confirms once more the relevance of
giving up local commutativity in the bulk. By means of the identity
\be
\lim_{d \downarrow 1}\, \frac{1}{\Gamma(d-1)}\, (p^2)_+^{d-2} = 
\delta (p^2) \, ,
\label{}
\end{equation}
which holds \cite{GS} in the sense of distributions, one recovers 
from (\ref{conf}) the free
massless field in the limit $d\to 1$. This fact explains the choice 
of normalization in
(\ref{confden}).

Summarizing, we described in this section the influence of a brane on the
bulk quantization. We considered for illustration the case
of $s=1$ noncompact extra dimensions, but the framework and the results
have a direct generalization to $s\geq 1$.
In that case $\lambda \mapsto (\lambda_1,...,\lambda_s)$,
where $\lambda_i$ will have discrete spectrum if the corresponding dimension is
compact.

\bigskip

\section{Currents induced on a brane}

The aim of this section is to display a link between bulk and brane
symmetries, establishing further interesting features of the class
$\Sr$. We begin with some preliminary  considerations on classical level.
Let $J_\alpha $ be a conserved current in the bulk.
With our choice of coordinates, this means that
\be
\der^\mu J_\mu (x,y) - \der_y J_4 (x,y) = 0 \, , \qquad \mu = 0,...,3 \, .
\label{cons}
\end{equation}
The corresponding charge
\be
Q = \int_{-\infty}^\infty d^3x \int_0^\infty dy J_0 (x,y) \, ,
\label{charge}
\end{equation}
is time-independent if $J_\alpha$ decay fast enough when both
$|\bx|\to \infty$, $y\to \infty$ and
\be
\int_{-\infty}^\infty d^3x J_4(x,0) = 0 \, .
\label{icbc}
\end{equation}
Eq.(\ref{icbc}), which is a direct consequence of (\ref{cons}) and the Gauss
divergence theorem, represents a kind of integral boundary condition 
with transparent
physical interpretation: $Q$ is conserved if and only if the total flux of
the bulk current across $\RR^3 \subset {\bf M}_{3+1}$ vanishes for any $x^0$.
Because of (\ref{cons}), the boundary induced current
\be
j_\mu (x) = \lim_{y \downarrow 0} J_\mu (x,y) \, ,
\label{ic}
\end{equation}
is not conserved in general. As suggested by eq.(\ref{cons}),
in order to construct a conserved current in ${\bf M}_{3+1}$, we
introduce the brane scalar field
\be
\chi (x) = \lim_{y \downarrow 0} \der_y J_4 (x,y) \, ,
\label{chi}
\end{equation}
and define
\be
k_\mu(x) = j_\mu (x) - \left (\der_\mu\, \Delta_0*\chi \right )(x) \, ,
\label{k}
\end{equation}
where $*$ denotes a convolution and $\Delta_0$ is defined by 
eq.(\ref{tf}). Now,
eqs.(\ref{tf},\ref{cons},\ref{k}) imply
\be
\der^\mu k_\mu (x) = 0 \, ,
\label{kcons}
\end{equation}
if
\be
\der^\mu \lim_{y \downarrow 0} J_\mu (x,y)  =
\lim_{y \downarrow 0} \der^\mu J_\mu (x,y) \, .
\label{cc}
\end{equation}
Together with (\ref{icbc}), the condition (\ref{cc}) is a basic
requirement for the validity of our construction below.

Summarizing these classical considerations, we have seen that a
conserved bulk current $J_\alpha$ induces both a vector current $j_\mu$ and
a scalar field $\chi$ on the brane $\der\M$. Generally, $j_\mu$ is 
not conserved.
The improved current $k_\mu$ is conserved, but is expected to have 
worse localization
properties, due to the convolution appearing in eq.(\ref{k}).

We turn now to the quantum case, focusing on
\be
S_0 = \int_{-\infty}^\infty d^4x  \int_0^\infty dy
:\left ( \der^\alpha \Phi^* \der_\alpha \Phi - M^2 \Phi^* \Phi \right ): (x,y)
  - \eta \int_{-\infty}^\infty d^4x :\Phi^* \Phi : (x,0) \, ,
\label{actc}
\end{equation}
which describes a free complex scalar field
\be
\Phi (x,y) = \frac{1}{\sqrt 2}\left [\Phi_1(x,y) + i \Phi_2(x,y)\right ] \, ,
\label{dec}
\end{equation}
whose real components $\Phi_1$ and $\Phi_2$ satisfy eqs.(\ref{eqm},\ref{bcphi}). The invariance of (\ref{actc}) under global $U(1)$ transf 
ormations implies the
conservation of the current
\be
J_\alpha (x,y) =
i: \left[\Phi^*(\der_\alpha \Phi ) - (\der_\alpha \Phi^*)\Phi \right 
] : (x,y) \, .
\label{sc}
\end{equation}
Due to the boundary condition (\ref{bcphi}), one has
\be
\lim_{y \downarrow 0} J_4(x,y) = 0 \, , \qquad \forall \, x \in {\bf 
M}_{3+1}\, ,
\label{J4}
\end{equation}
which shows in turn that the current (\ref{sc}) satisfies (\ref{icbc}). Though
$J_4$ vanishes on the brane, we emphasize that its $y$-derivative 
$\chi$ does not.

{}Fixing the data $\simp \in \Si$, we quantize both $\Phi_1$ and 
$\Phi_2$ according to
the scheme developed in the previous section and are in position to compute any
correlation function of the operators $J_\alpha , \Phi^*, \Phi $. The latter
induce on the brane the fields $j_\mu, \chi, \varphi^*, \varphi $ by means of
\be
\qquad \qquad \langle j_{\mu_1} (x_1)\cdots \chi (x_k)\cdots
\varphi^* (x_{k+m+1})\cdots \varphi (x_{k+m+n+1})\cdots  \rangle_{{}_0} =
\qquad \qquad \nb
\end{equation}
\be
\lim_{y_i \downarrow 0} \langle J_{\mu_1}(x_1,y_1)\cdots 
\der_{y_k}J_4 (x_k,y_k) \cdots
\Phi^* (x_{k+m+1},y_{k+m+1})\cdots \Phi (x_{k+m+n+1},y_{k+m+n+1})
\cdots \rangle_{{}_0} \, .
\label{corg1}
\end{equation}
It is clear from eq.(\ref{corg1}) that $\varphi$ inherits the 
$U(1)$-symmetry from $\Phi$,
which poses two main questions:

\begin{description}

\item {(a)} does a conserved current, generating the brane
$U(1)$-invariance, exist?

\item {(b)} in case of affirmative answer to (a), does this current
satisfy a Ward identity?

\end{description}
 
\noindent The answers to (a) and (b) are encoded in the correlation functions
(\ref{corg1}). Let us consider the relevant ones, starting by
\bea
\langle j_\mu (x_1)j_\nu (x_2)\rangle_{{}_0} =
\qquad \qquad \qquad \qquad \qquad \qquad \qquad \nb \\
2\int_0^\infty d\lambda^2_1\, d\lambda^2_2 \,
\varrho (\lambda^2_1 )\varrho (\lambda^2_2 )
\left [ (\der_\mu W_{\lambda^2_1})(x_{12})
(\der_\nu W_{\lambda^2_2})(x_{12}) -
(\der_\mu \der_\nu W_{\lambda^2_1})(x_{12}) W_{\lambda^2_2}(x_{12})\right ]
\, .
\label{bcc}
\ena
Notice that the distribution product at coinciding points in the
integrand of eq. (\ref{bcc}) is  well-defined and simply related to
the product $W_{\lambda^2_1}(x) W_{\lambda^2_2}(x)$, whose Fourier transform
\bea
\int_{-\infty}^\infty \frac {d^4p}{(2\pi)^4} \,
\e^{-ipx} W_{\lambda^2_1}(x) W_{\lambda^2_2}(x) =
\qquad \qquad \qquad \nb \\
\frac{1}{8\pi p^2}\, \theta (p^0) \theta (p^2-(\lambda_1+\lambda_2)^2)
\sqrt{p^2-(\lambda_1+\lambda_2)^2}\,
\sqrt{p^2-(\lambda_1-\lambda_2)^2}
\label{phasespace}
\ena
represents the phase space of two relativistic scalar particles of mass
$\lambda^2_1$ and $\lambda^2_2$. Eq.(\ref{bcc}) implies
\be
\langle \der^\mu j_\mu (x_1)\der^\nu j_\nu (x_2)\rangle_{{}_0} =
\int_0^\infty d\lambda^2_1\, d\lambda^2_2 \,
\varrho (\lambda^2_1 )\varrho (\lambda^2_2 )\, (\lambda_1^2-\lambda_2^2)^2 \,
W_{\lambda^2_1}(x_{12}) W_{\lambda^2_2}(x_{12}) \, ,
\label{bdd}
\end{equation}
which confirms that $j_\mu$ is not conserved. In fact, one can deduce from the
correlators (\ref{corg1}) that
\be
\der^\mu j_\mu (x) = \chi (x) \, .
\label{jx}
\end{equation}
Therefore, the improved quantum current $k_\mu$ is
conserved. In order to demonstrate that it is precisely $k_\mu$ which 
generates the
brane $U(1)$ symmetry, one must verify the corresponding Ward 
identities. For this
purpose we consider the vertex functions
\bea
\langle T j_\mu (x_1)\varphi^* (x_2)\varphi (x_3) \rangle_{{}_0} =
\qquad \qquad \qquad \qquad \qquad \nb \\
i\int_0^\infty d\lambda^2_1\, d\lambda^2_2 \,
\varrho (\lambda^2_1 )\varrho (\lambda^2_2 )
\left [ (\der_\mu \Delta_{\lambda^2_1})(x_{12}) \Delta_{\lambda^2_2}(x_{13}) -
\Delta_{\lambda^2_1}(x_{12}) (\der_\mu \Delta_{\lambda^2_2})(x_{13})\right ]
\, ,
\label{bcpp}
\ena
\be
\langle T \chi (x_1)\varphi^* (x_2)\varphi (x_3) \rangle_{{}_0} =
i\int_0^\infty d\lambda^2_1\, d\lambda^2_2 \,
\varrho (\lambda^2_1 )\varrho (\lambda^2_2 )\, (\lambda_2^2-\lambda_1^2)\,
\Delta_{\lambda^2_1}(x_{12}) \Delta_{\lambda^2_2}(x_{13})
\, .
\label{xpp}
\end{equation}
Combining eqs.(\ref{jx}-\ref{xpp}) one gets
\be
\der^\mu\langle T k_\mu (x_1)\varphi^* (x_2)\varphi (x_3) \rangle_{{}_0} =
iC\delta(x_{12})\langle T \varphi^* (x_2)\varphi (x_3) \rangle_{{}_0} -
iC\delta(x_{13})\langle T \varphi^* (x_3)\varphi (x_2) \rangle_{{}_0}
\, ,
\label{wi}
\end{equation}
which is the conventional Ward identity, provided that the bound 
(\ref{ffr}) holds.
Otherwise, the right hand side of (\ref{wi}) diverges, which demonstrates the
special status of the brane fields corresponding to $\Sr$. We 
emphasize that the above
analysis takes into account not only the zero mode, but the whole KK-tower
as well. The problem of Ward identities in the AdS/CFT framework has 
been faced
in \cite{C}.

In conclusion, a conserved current $J_\alpha $ in the bulk induces both a
vector current $j_\mu$ and scalar $\chi$, the latter being essential in
constructing a conserved brane current. We expect therefore the $\chi$-like
degrees of freedom to be fundamental in model building with extra 
space dimensions.

\bigskip

\section{\bf Anti-de Sitter bulk space}

In this section we keep the bulk manifold $\M = \RR^4 \times \RR_+$,
but now equipped with the AdS metric
\be
G_{\alpha \beta } = \begin{pmatrix} g\, \e^{-2ay} & 0 \\ 0 & 
-1\end{pmatrix}\, ,
\qquad a>0 \, .
\label{met}
\end{equation}
Thus $\M$ is a slice of the five-dimensional AdS space-time, whose
boundary $\der \M$ still coincides with ${\bf M}_{3+1}$.
The problem is to construct and investigate the scalar quantum field $\Phi $
defined by
\be
S_0 = \frac{1}{2} \int_{-\infty}^\infty d^4x  \int_0^\infty dy\, \sqrt {G}
:\left ( \der^\alpha \Phi \der_\alpha \Phi - M^2 \Phi \Phi \right ): (x,y)
  - \frac{\eta }{2}\int_{-\infty}^\infty d^4x \sqrt {g} :\Phi \Phi : (x,0) \, .
\label{actA}
\end{equation}
{}From (\ref{actA}) one infers the equation of motion
\be
\left (\der^\mu \der_\mu - \e^{2ay}\der_y\, \e^{-4ay} \der_y + 
\e^{-2ay} M^2 \right )
\Phi (x,y) = 0 \, ,
\label{eqmA}
\end{equation}
and the boundary condition (\ref{bcphi}). The quantization of $\Phi$ follows
the scheme developed in section 2. A modification of the
functional input (i) is required by the fact that instead of the 
Hamiltonian $-\der_y^2$,
one has now
\be
-\e^{2ay}\der_y\, \e^{-4ay} \der_y + \e^{-2ay} M^2 \, .
\label{y}
\end{equation}
The operator (\ref{y}) defines a well-known \cite{T,N} singular boundary value
problem on $\RR_+$, related to Bessel's equation. Setting
\be
\nu = \sqrt {4 + \frac{M^2}{a^2}} \, ,  \qquad
\eta_b = (2-\nu) a \, ,
\label{ni}
\end{equation}
one has \cite{MP}:
\be
\psi (y,\lambda) = \e^{2ay}
\left [J_\nu (\lambda a^{-1} \e^{ay})
{\widetilde Y}_\nu (\lambda a^{-1}) -
Y_\nu(\lambda a^{-1} \e^{ay})
{\widetilde J}_\nu (\lambda a^{-1})\right ] \, , \qquad \lambda \in \RR_+\, ,
\label{bess}
\end{equation}
\be
\psi_b(y) = \delta_{\eta \eta_b}\, \sqrt {2a(\nu -1)}\, \e^{\eta_by}\, ,
\label{bs1}
\end{equation}
with
\be
{\widetilde Z}_\nu (\zeta ) = \frac{1}{2\sqrt{1+\widetilde{\eta}^2}}
\left[(1-2\widetilde {\eta})Z_\nu(\zeta ) + 2\zeta Z_\nu^\prime (\zeta )
\right ] \, ,
\qquad \widetilde {\eta} = \frac{\eta}{a} - \frac{3}{2} \, ,
\label{til}
\end{equation}
$Z_\nu$ denoting the Bessel function $J_\nu$ or $Y_\nu$ of first and
second kind respectively. Instead of (\ref{compl}), now the 
completeness relation is
\be
\int_0^\infty d\lambda \, \mu (\lambda )\,
{\overline \psi} (y_1,\lambda) \psi (y_2,\lambda) +
\delta_{\eta \eta_b}\, \psi_b(y_1)\, \psi_b(y_2) =
\e^{2ay_1}\, \delta (y_{12}) \, ,
\label{compl1}
\end{equation}
where
\be
\mu (\lambda ) =
\frac{\lambda a^{-1}}{{\widetilde J}_\nu^2 (\lambda a^{-1}) +
{\widetilde Y}_\nu^2 (\lambda a^{-1})} \, .
\label{mesures}
\end{equation}
The algebraic input (ii) must be also slightly modified: the creation
and annihilation operators
$\{a^*(\bp,\lambda ),\, a(\bp,\lambda ),\, b^*(\bp),\, b(\bp)\}$
satisfy eqs.(\ref{ccr1}-\ref{ccr00}) with the replacement
\be
2\pi \delta (\lambda_1-\lambda_2) \longmapsto
\frac{1}{\mu (\lambda_1)}\, \delta (\lambda_1-\lambda_2) \, .
\label{repl}
\end{equation}
The quantum field $\Phi$ in the AdS background is then given by 
eq.(\ref{f1}) with
the substitutions $d\lambda \mapsto 2\pi\, \mu (\lambda )\, d\lambda $ and
(\ref{bess},\ref{bs1}). The field $\varphi$ induced
on the brane is fully determined by its two-point function
\be
w(x_{12}) = \langle \varphi (x_1)\varphi (x_2)\rangle_{{}_0} =
\int_0^\infty d\lambda^2 \, \varrho_{{}_{\rm AdS}} (\lambda^2 )\, 
W_{\lambda^2}(x_{12}) \, ,
\label{corAdS}
\end{equation}
where
\be
\varrho_{{}_{\rm AdS}} (\lambda^2) =
\theta (\lambda^2) \,
\frac{2\mu (\sqrt {\lambda^2}\, )\, \sigma (\sqrt {\lambda^2}\, )}
{\pi^2\, (1+\widetilde{\eta}^2 )\, \sqrt {\lambda^2}}
+ 2\sigma_b \delta_{\eta\eta_b}\, a(\nu -1) \delta(\lambda^2) \, .
\label{rhoAdS}
\end{equation}
As expected, the K\"all\'en-Lehmann measure $\varrho_{{}_{\rm AdS}} 
(\lambda^2 )d\lambda^2$
keeps trace of $\simp \in \Si$. Due to the interaction of $\Phi$ with 
the AdS background,
there is in general no mass gap on the brane even for $M\not=0$. 
Bearing in mind this
novelty with respect to the flat case, the results of section 2 apply 
also to the slice of
AdS space-time, considered above.

\bigskip

\section{Conclusions}

We have studied above quantum field theory on a 4+1 dimensional bulk 
manifold with
boundary - a 3-brane which represents the observable 3+1 dimensional
space-time. The brane breaks the bulk Lorentz
symmetry down to brane Lorentz symmetry and allows for nonlocal initial
conditions compatible with the bulk symmetries.
Technically, these conditions are
parametrized by $\simp \in \Si$. Each pair $\simp $ defines
a generalized canonical commutation structure and provides a direct relation
between the initial conditions for the time evolution in the bulk and
the KK spectral measure on the brane. We have constructed the
bulk fields associated with $\Si $ and their brane relatives,
whose basic properties are captured
by the measure of the K\"all\'en-Lehmann representation of the
two-point function. From the latter we have inferred the existence of a
subset $\Sr \subset \Si$, generating renormalizable models on the brane.
In this context we have investigated also the interplay between 
conserved bulk and brane
currents and the validity of brane Ward identities. Our analysis 
covers the full
spectrum of bulk excitations and sheds new light on the correspondence
$\{\varphi_i\} \leftrightarrow \{\Phi_i\}$ between brane and bulk
quantum fields, showing that nonlocal theories in $\M$ may induce local
theories on $\der \M$. The results, obtained initially in flat
bulk space, are extended to the case of AdS background as well.

The freedom in the choice of initial conditions for the
bulk quantization, discussed in this work, is an universal intrinsic feature
of quantum field theory with extra dimensions, which must be taken in
consideration. It can be used for improving the UV behavior of brane 
interactions
and for inducing conformal covariant fields on the brane. We expect that
nonlocal bulk fields will help for avoiding some no-go theorems 
\cite{MN} for the
construction of Randall-Sundrum compactifications from supergravity.
At the same time, the possibility to adopt various initial conditions
$\simp \in \Si$ reflects an
ambiguity, which is not fixed by quantum field theory itself.
In order to obtain some information about $\simp $,
one has to resort most probably to a more fundamental
setting like string theory. Clarifying this
point represents an interesting open problem for further investigations.

\bigskip
\bigskip


\end{document}